\documentclass[utf8]{frontiersinFPHY_FAMS2} % Vancouver Reference Style (Numbered) for articles in the journals "Frontiers in Physics" and "Frontiers in Applied Mathematics and Statistics" 

\setcitestyle{square} % for articles in the journals "Frontiers in Physics" and "Frontiers in Applied Mathematics and Statistics" 
\usepackage{url,hyperref,lineno,microtype,subcaption}
\usepackage{mathrsfs}
\usepackage[onehalfspacing]{setspace}

\def\keyFont{\fontsize{8}{11}\helveticabold }
\def\firstAuthorLast{}
\def\Authors{J\'er\'emy Besson$^{1,2}$, Javier Carballo$^{3}$, Christiana Pantelidou$^{4,*}$ and Benjamin Withers$^{3,*}$}
\def\Address{$^{1}$Institut  de  Math\'ematiques  de  Bourgogne  (IMB),  CNRS, Universit\'e  de  Bourgogne,  Dijon,  France\\
$^{2}$Albert-Einstein-Institut, Max-Planck-Institut für Gravitationsphysik, Hannover, Germany and Leibniz Universit\"at Hannover, Hannover, Germany\\
$^{3}$Mathematical Sciences and STAG Research Centre, University of Southampton, Southampton, United Kingdom\\
$^{4}$School of Mathematics and Statistics, University College Dublin, Dublin, Ireland}
\def\corrAuthor{Christiana Pantelidou}

\def\corrEmail{christiana.pantelidou@ucd.ie}

\newcommand{\bea}{\begin{eqnarray}}
\newcommand{\eea}{\end{eqnarray}}
\newcommand{\be}{\begin{equation}}
\newcommand{\ee}{\end{equation}}
\newcommand{\ba}{\begin{align}}
\newcommand{\ea}{\end{align}}

\newcommand{\im}{\mathfrak{Im}\,}

\usepackage{cprotect}
\begin{document}
\onecolumn
\firstpage{1}
% \pagenumbering{arabic}

\title[Transients in black hole perturbation theory]{Transients in black hole perturbation theory} 

\author[\firstAuthorLast ]{\Authors} %This field will be automatically populated
\address{} %This field will be automatically populated
\correspondance{} %This field will be automatically populated

%\extraAuth{}% If there are more than 1 corresponding author, comment this line and uncomment the next one.
\extraAuth{Benjamin Withers\\b.s.withers@soton.ac.uk}

\maketitle

\begin{abstract}
Black hole quasinormal modes arise as eigenmodes of a non-normal Hamiltonian and consequently they do not obey orthogonality relations with respect to commonly used inner products, for example, the energy inner product. A direct consequence of this is the appearance of transient phenomena. This review summarises current developments on the topic, both in  frequency- and time-domain. In particular, we discuss the appearance of i) transient plateaus: arbitrarily long-lived sums of quasinormal modes, corresponding to localised energy packets near the future horizon; ii) transient growth, with the latter either appearing in the vicinity of black hole phase transitions or in the context of higher-derivative Sobolev norms. 
\section{}

\tiny
 \keyFont{ \section{Keywords:} non-modal, quasinormal modes, black holes, transients, pseudospectra, black hole spectroscopy, non-normal, ringdown} 
\end{abstract}

\section{Introduction} 

An indispensable tool in the study and characterisation of the dynamics of black holes is their spectrum of quasinormal modes (QNMs) -- for recent reviews see~\cite{Berti:2009kk,Konoplya:2011qq}. QNMs are solutions to the wave equation arising when general relativity is considered perturbatively at linear order, and they determine how small perturbations evolve over time,  capturing their ‘ringdown’ behaviour.\footnote{Second order QNMs, usually referred to as QQNMs, have also been constructed recently~\cite{Lagos:2022otp,Pantelidou:2022ftm}.} As such, QNMs have received a lot of attention in the literature. Within holography, they determine the near-equilibrium properties of strongly coupled quantum field theories, in particular some transport coefficients, such as viscosity, conductivity and diffusion constants~\cite{Policastro:2001yc,Kovtun:2005ev}. In astrophysics, the detection of QNMs in gravitational wave experiments would allow precise measurements of the mass and spin of black holes -- through the so-called black hole spectroscopy programme~\cite{Baibhav:2023clw} -- as well as new tests of general relativity. Similarly, QNMs also serve as indicators of black hole instabilities: a single unstable mode signals exponentially growing perturbations leading to a new equilibrium configuration, which is particularly important in higher dimensions as well as in the holographic context. In addition, QNMs also play an instrumental role in semiclassical gravity, e.g. in the context of Hawking radiation~\cite{York:1983zb}, as well as in Mathematical Relativity, e.g. in understanding properties of Cauchy horizons~\cite{Hintz:2015jkj}.

The defining property of a black hole is its event horizon, through which energy dissipates. This dissipative nature of black holes has a direct imprint on the operator that gives rise to  QNMs: the operator is non-normal. This absence of normality leads to the QNM eigenfunctions being neither orthogonal\footnote{With respect to standard choices of inner product. See \cite{Jafferis:2013qia, Green:2022htq, London:2023aeo, Arnaudo:2025bnm} for the construction of QNM orthogonality relations in other products.} nor complete, while the QNM frequencies are highly sensitive to small perturbations, resulting in spectral instability. These features substantially complicate the interpretation of QNMs and, in fact, in certain contexts question the validity of their use. Note that non-normality is a generic feature of dissipative systems and as such, has been observed and investigated in both (i) quantum mechanics, where the introduction of non-selfadjoint operators in PT-symmetric quantum mechanics entails that the associated spectrum is insufficient to draw full, quantum-mechanically relevant conclusions \cite{Krejcirik:2014kaa}, and in (ii) fluid dynamics in relation to the transition between laminar and turbulent flows \cite{trefethen2005spectra}. 

In essence, to-date, we have only explored the `tip of the iceberg' in terms of non-normality in black hole physics, especially in dynamical settings, where the non-orthogonality of QNMs can give rise to short-term, transient phenomena. Here we review progress in this direction.

In order to set the stage, in what follows we foliate spacetime with \emph{hyperboloidal slices}, $\Sigma_\tau$ -- spacelike slices that pierce the future event horizon. These slices are labeled by time $\tau$ and are traversed by a radial coordinate $z$ with the future event horizon reached at $z=1$. For brevity we suppress dependence in the transverse directions. In the spacetimes we consider here, the equation of motion for a perturbation $\psi(\tau, z)$ (scalar, electromagnetic, or spin-2), will obey a first-order-reduced equation of motion,
\begin{equation}
i\partial_\tau u=\mathcal{H} u,
\label{eq:hamiltonian}
\end{equation}
where $\mathcal{H}$ is a 2$\times$2 matrix and a second-order differential operator in $z$ and $u=(\psi, \partial_\tau \psi)^T$. For initial data $u(0, z)$, the time-dependent solution  of the system is given formally as $u(\tau, z) = e^{-i\mathcal{H}\tau} u(0, z)$, in terms of the evolution operator $e^{-i\mathcal{H}\tau}$.  Given a harmonic decomposition $u(\tau,z) \sim\chi(z) e^{-i \omega \tau}$, QNMs are defined as solutions to the eigenvalue problem 
\begin{equation}
\omega_n \chi_n=\mathcal{H} \chi_n,
\end{equation}
subject to ingoing behaviour at the future event horizon and appropriate boundary conditions at infinity. Then, the spectrum of the theory is given by $\sigma(\mathcal{H})=\{\omega_n, n\ge0\}$. We can define an energy associated to matter on a hyperboloidal slice by
\begin{equation}
E \equiv \int_{\Sigma_\tau}T^{\mu}_{\phantom{z}\tau}n_{\mu}\,d\Sigma_\tau,\label{Edef}
\end{equation}
where $n=\frac{-1}{\sqrt{-g^{\tau\tau}}}d\tau$ is the unit, future-directed normal to $\Sigma_\tau$. $T^{\mu\nu}$ is the matter stress-energy tensor, and is at least quadratic in the perturbation $\psi$. Note that $T^{\mu\nu}$ can contain contributions from other fields. Due to local conservation of the currents $T^{\mu\nu}$, the total energy $E$ is conserved up to boundary terms.
The energy of $\psi$ on $\Sigma_\tau$ is then given by \eqref{Edef} with $T_{\mu\nu}=T_{\mu\nu}^\psi$, from which the \emph{energy inner product} $\langle\cdot,\cdot\rangle_E$ (see \cite{Gasperin:2021kfv} for an extended discussion) is defined such that
\be
E[u]=\langle u,u\rangle_E = \|u\|_E^2.
\label{eq:energy_norm}
\ee

\section{Insights from the pseudospectrum}\label{section2}
One can extract various insights about the time domain problem from spectral features. In particular, a useful object is the pseudospectrum,
\begin{equation} 
\sigma_\epsilon(\mathcal{H})=  \left\{\omega\in\mathbb{C}\, \big|\,\omega\in\sigma(\mathcal{H}+ \delta \mathcal{H}), ||\delta \mathcal{H}||\le \epsilon\right\},\label{pseudospectrum}
\end{equation}
which, along with many of the definitions in this section, can be found in \cite{trefethen2005spectra}.
In the black hole context, \eqref{pseudospectrum} has received much attention as a way to assess the stability of QNM frequencies under environmental perturbations \cite{Jaramillo:2020tuu}, building upon the seminal observations of \cite{Nollert:1998ys,Nollert:1996rf}. Heuristically, $\sigma_\epsilon$ at fixed $\epsilon$ provides the contours of a useful topographic map of the complex frequency plane. Peaks are infinitely high and correspond to the point spectrum, while the width of the peaks have something to say about the associated spectral stability properties and transient effects.

In particular, for our purposes, a significant protrusion of pseudospectral contour lines into the unstable-half $\omega$-plane points towards transient phenomena (in our conventions this is the upper-half $\omega$-plane). A lower bound on the peak growth of the evolution operator is given as follows,
\begin{equation}\label{eq:lowerBound}
     \sup_{\tau\geq0} ||e^{-i  \mathcal{H} \tau}||\ge \frac{\alpha_\epsilon(\mathcal{H})}{\epsilon}\,,\quad \forall \epsilon>0\,,
 \end{equation}
where we have introduced the pseudospectral abscissa, $\alpha_\epsilon(\mathcal{H})=\sup \, \im \sigma_\epsilon(\mathcal{H})$. The strongest lower bound is given by the Kreiss constant, $\mathcal{K}(\mathcal{H})=\sup_{\epsilon > 0} \alpha_\epsilon(\mathcal{H})/\epsilon$. Relatedly, an upper bound on growth follows from
\be
|| e^{-i  \mathcal{H} \tau}||\le e^{-i \mathfrak{w}(\mathcal{H})\tau}, \qquad \forall \tau\ge0,
\ee
where we have introduced the numerical abscissa $\mathfrak{w}(\mathcal{H})=\sup_{\epsilon>0} (\alpha_\epsilon(\mathcal{H})-\epsilon)$.

In the black hole context, these quantities were first studied in~\cite{Jaramillo:2022kuv} in the context of binary black hole mergers in the close-limit approximation.\footnote{See also \cite{Boyanov:2022ark} for a related study of extreme compact objects, where a Kreiss constant consistent with $K(\mathcal{H}) = 1$ was obtained numerically by computing the ratio of the pseudospectral abscissa $\alpha_\epsilon(\mathcal{H})/\epsilon$ in the limit $\epsilon\to\infty$.} Specifically, in the case of a Schwarzschild black hole in the energy norm \eqref{eq:energy_norm}, \cite{Jaramillo:2022kuv} computed the numerical abscissa to be $\mathfrak{w}(\mathcal{H})=0$, which implies $\mathcal{K}(\mathcal{H})=1$. This implies there is no growth of energy of a perturbation in the exterior of Schwarzschild spacetime, which is simply a consequence of energy conservation \cite{Carballo:2024kbk}.\footnote{Note that \cite{Chen:2024mon} 
reports transient growth in the context of Kaluza-Klein black holes in Gauss-Bonnet gravity. However, the system studied in \cite{Chen:2024mon} is conservative up to boundary terms and (3.19) there can be written as a total derivative. As such, the reported result on transient growth is incorrect.}

Going further, one may ask if the pseudospectrum can be used to identify scenarios in which perturbations of black holes can grow. However, a critical issue arises when \eqref{pseudospectrum} is considered more generally in the black hole context. This is most easily stated using the following equivalent definition of \eqref{pseudospectrum}, which utilises the norm of the resolvent,
\begin{equation} 
\sigma_\epsilon(\mathcal{H})=  \left\{\omega\in\mathbb{C}\, \big|\,\Vert{R_\mathcal{H}(\omega)}\Vert=\Vert{(\mathcal{H}-\omega I)^{-1}}\Vert\geq1/\epsilon\right\}.\label{pseudospectrum_resolvent}
\end{equation}
When the resolvent operator is approximated as a matrix for the purposes of numerical evaluation it does not always converge with increasing resolution \cite{Boyanov:2024}. See \cite{warnick2024stability,BessonV2} for further discussions. However, it is proven in \cite{Warnick:2013hba} for asymptotically AdS and dS black holes that the norm of the resolvent exists in a band structure in the complex $\omega$ plane provided one uses a particular class of higher-derivative norms. There, higher-derivative norms were introduced in order to impose a higher degree of regularity for the purposes of defining QNMs.
This motivates the use of higher-derivative norms both in the evaluation of pseudospectra and for assessing transient phenomena, as in \cite{BessonV2}. 
In \cite{BessonV2} the following higher-derivative norms are defined\footnote{Note that this is different to the corresponding inner product used in \cite{Boyanov:2024}.}
\be
\langle u_1, u_2\rangle_{_{H^p}} 
= \sum_{j=0}^p \langle\partial_x^j u_1,
\partial_x^j u_2\rangle_{E} \,, \label{sobolevdef}
\ee
referred to as the Sobolev $H^p$-inner product, where here $(\tau,x)$ refer to the Bizoń-Mach hyperboloidal coordinates \cite{Bizon:2020qnd} for the P\"oschl-Teller model. Note $p=0$ corresponds to the energy-norm.

The Kreiss constant was also discussed in~\cite{Carballo:2025ajx}, where it was extracted from the pseudospectrum of a truncated Hamiltonian, $\mathcal{H}_W$, where the functional space was restricted to a subspace $W$ of the first $M$ quasinormal modes. \cite{Carballo:2025ajx} found that $\mathcal{K}(\mathcal{H}_W)>1$, for a system describing charged scalar perturbations in a Reissner-Nordström (RN) - AdS$_4$ black brane. This indicates that there exist perturbations that exhibit transient energy growth in the scalar field when all the modes are stable.

\section{Time domain}
In the last section, we presented quantities computed from the pseudospectrum (and its respective limits) that provide insights into the time evolution of linear perturbations. In particular, a non-zero numerical abscissa, $\mathfrak{w}(\mathcal{H})>0$, and thus $\mathcal{K}(\mathcal{H})>1$, immediately implies that there exist perturbations whose time evolution exhibit transient growth in the observable defined by the chosen norm $\|\cdot\|$. The pseudospectral analysis is however incomplete, since it does not capture important transient effects that arise even in the absence of growth \cite{Carballo:2024kbk}, and should be complemented with a full time domain evolution of perturbations. 

Consider a black hole coupled to a scalar field. A natural choice of observable is the energy of the scalar field $\psi$ on hyperboloidal slices as given by the energy norm $\|\cdot\|_E$ \eqref{eq:energy_norm}. A key feature now is that energy dissipation through the horizon and to $\mathscr{I}^+$ renders a non-normal $\mathcal{H}$ in \eqref{eq:hamiltonian} under $\langle\cdot,\cdot\rangle_E$, and thus its regular normalisable eigenfunctions (the QNMs) are not orthogonal under this product. Consequently, the energy of a perturbation formed from a sum of QNMs, $u(\tau,z)=\sum_{n=1}^{M}c_ne^{-i\omega_n\tau}\chi_n(z)$, is not just the sum of the energies of each individual QNM, but rather
\be
E[u]=\sum_{n=1}^{M}|c_n|^2 e^{2\im{\omega_n}\tau} E[\chi_n] + \text{cross-terms}.
\label{eq:cross_terms}
\ee
There are cross-terms arising from the non-orthogonality of QNMs under \eqref{eq:energy_norm} that allow for non-trivial transient dynamics. Note that without the cross-terms, the slowest possible energy decay is set by the fundamental mode $\omega_0$.

In this context, the first systematic time domain study of transients in black hole perturbations was introduced in \cite{Carballo:2024kbk} using the \emph{energy growth curve}, $G(\tau)\equiv\|e^{-i\mathcal{H}\tau}\|_E^2$, and \emph{optimal perturbations} -- tools inherited from hydrodynamics \cite{Reddy93, Gustavsson_1991, Henningson_Lundbladh_Johansson_1993, ButlerFarrell, Reddy_Henningson_1993, Trefethen1993}. Considering a subspace of solutions $W$ to \eqref{eq:hamiltonian} spanned by the first $M=\dim(W)$ QNMs, $\{\chi_n\}_{n=1}^M$ (ordered by decreasing $\im{\omega}$), $G_W(\tau)\equiv \|e^{-i\mathcal{H}_W\tau}\|_E^2$ determines the maximum possible energy at a specific time $\tau$, relative to the energy at a fiducial intial time $\tau=0$, over all solutions in $W$. Optimal perturbations, $u_{\text{opt.}}\in W$, are then those that maximise the energy at a target time $\tau_*$ such that $E[u_{\text{opt.}}(\tau_*,z)]=G_W(\tau_*)$. Both the value of $G_W(\tau_*)$ and the set of coefficients $\vec{c}$ in the initial data expansion
\be
u_{\text{opt.}}(0,z)=\sum_{n=1}^Mc_n \chi_n(z)
\label{eq:initdata}
\ee
where QNMs are normalised $\langle \chi_n(z),\chi_n(z)\rangle_E=1$, are obtained from the singular value decomposition of $e^{-i H_W\tau_*}$ -- $H_W$ is the representation of $\mathcal{H}_W$ in an orthonormal basis of functions for $W$, an $M\times M$ matrix encoding the information of the spectrum (see \cite{Carballo:2024kbk, Carballo:2025ajx} for more details). Finally, $u_{\text{opt.}}$ evolves simply according to the time evolution of each QNM in \eqref{eq:initdata}, i.e.
\be
u_{\text{opt.}}(\tau,z)=\sum_{n=1}^Mc_n e^{-i \omega_n \tau}\chi_n(z).
\ee

\begin{figure}
\centering
\includegraphics[width=0.7\textwidth]{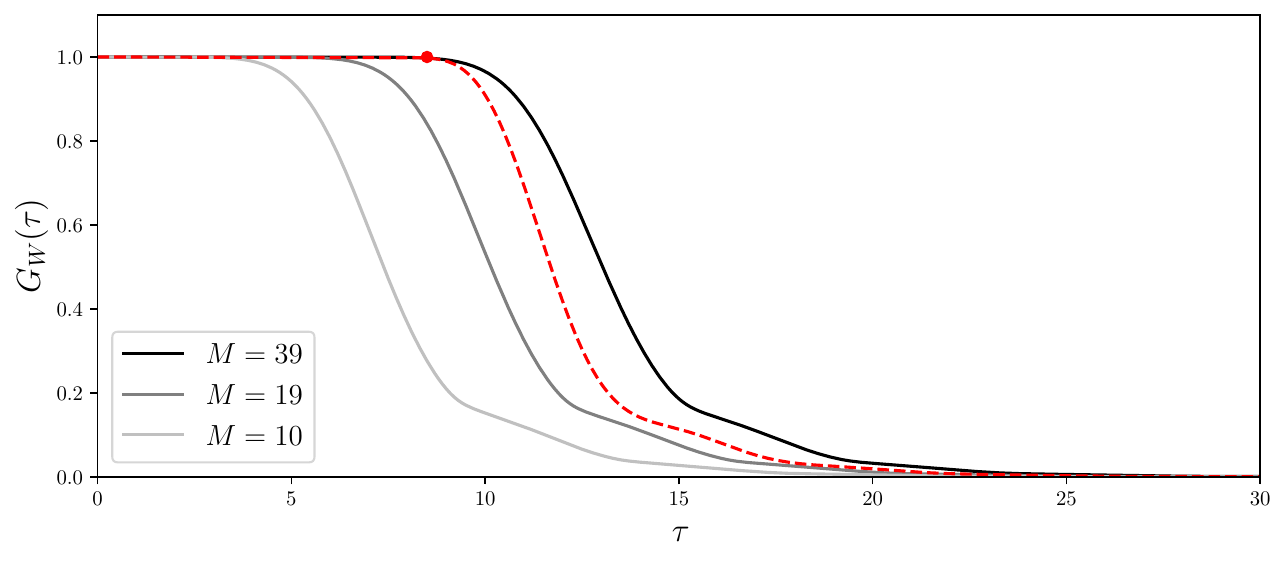}\\
\includegraphics[width=0.8\textwidth]{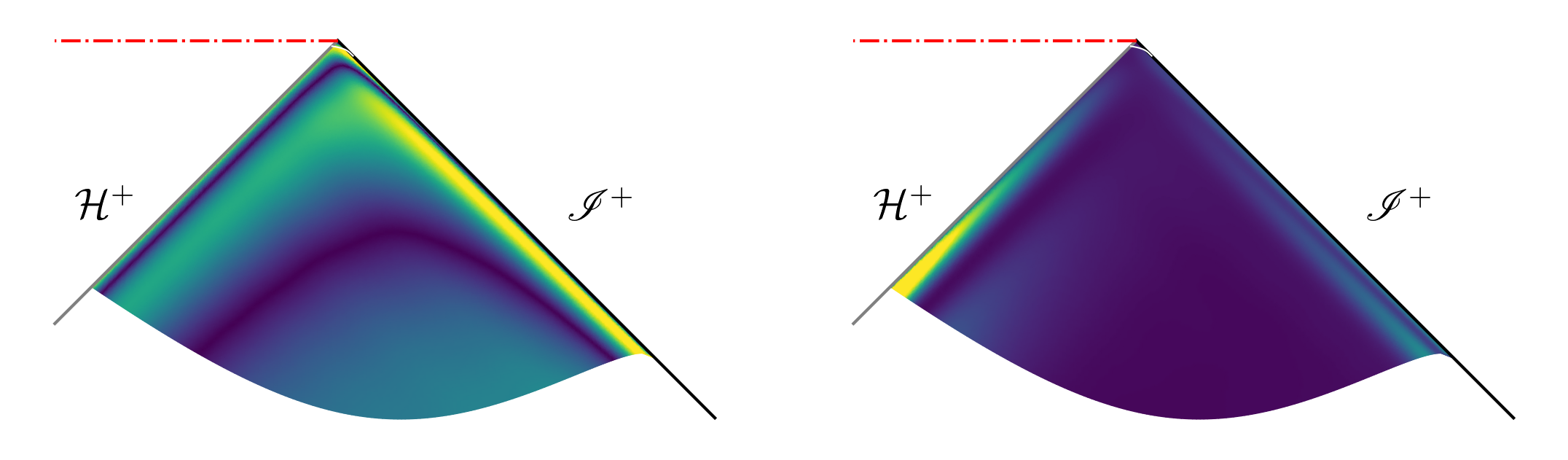}
    \caption{Energy growth curves and optimal perturbation for Schwarzschild $s=2,l=2$ Regge-Wheeler perturbations (figure taken from \cite{Carballo:2024kbk}). \textbf{Top:} $G_W$ for various $M=\dim(W)$ (solid curves), and the energy of an optimal perturbation of $M=39$ QNMs with $\tau_*=8.5$ (red-dash). \textbf{Bottom:} Modulus (left) and energy density (right) of the optimal perturbation in the conformal diagram of Schwarzschild. The energy is initially localised at $\mathcal{H}^+$ and $\mathscr{I}^+$, and then propagates along them until it starts dispersing and decaying at $\tau \simeq \tau_*$ (indicated by the white slice near $i^+$). The dash-dotted red line represents the curvature singularity.}
    \label{fig:schwarzschild}
\end{figure}

 Using this methodology, the main result of \cite{Carballo:2024kbk} consisted in demonstrating the existence and constructing (both analytically and numerically) arbitrarily long-lived linear black hole perturbations in a variety of spacetimes, due to transient effects, despite a lack of energy growth. An example of such perturbations for $s=2, l=2$ (spin and angular momentum) Regge-Wheeler QNMs of the Schwazschild black hole is presented in figure \ref{fig:schwarzschild}. The top panel shows $G_W$ for different values of $M$ indicating the absence of growth in the energy of perturbations, in accordance with the $\omega(\mathcal{H})=0$ and $\mathcal{K}(\mathcal{H})=1$ results discussed in section \ref{section2}. Note that the total (quadratic in perturbations) energy of the system $E$ can only stay constant or decay, and its only contribution is the energy of the gravitational perturbation. However, $G_W$ exhibits an initial transient plateau with duration $\sim\log M$ that demonstrates the existence of optimal perturbations with lifetimes scaling as $\log M$, followed by an exponential decay with the fundamental mode $\omega_0$ decay rate. The energy of such a perturbation, $E[u_{\text{opt.}}(\tau,z)]$, is displayed in red-dash. In the bottom panels, $|u_{\text{opt.}}|$ (left) and its energy density (right) are plotted in the conformal diagram. From the energy density, it is clear that $u_{\text{opt.}}$ is physically realised as localised energy packets travelling along $\mathcal{H}^+$ and $\mathscr{I}^+$ that do not either fall into the black hole or escape to infinity, respectively, until $\tau \simeq\tau_*$. Mathematically, this is a direct consequence of the non-orthogonality of QNMs under the energy norm \eqref{eq:energy_norm}, ultimately due to non-normality of $\mathcal{H}_W$, which leads to the cross-terms in \eqref{eq:cross_terms} allowing for cancellations in the sum that keep the energy constant.

\begin{figure}
\centering
\includegraphics[width=0.41\textwidth]{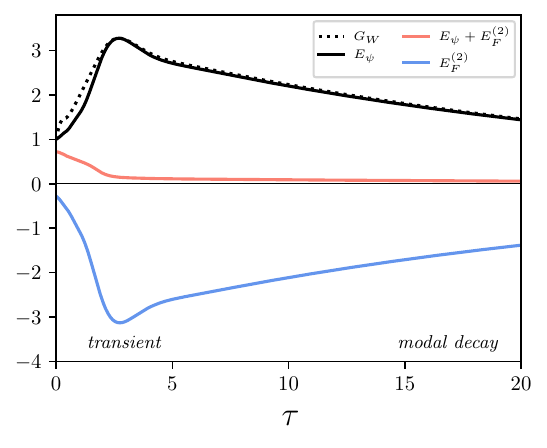}
\raisebox{0.35cm}{\includegraphics[width=0.52\textwidth]{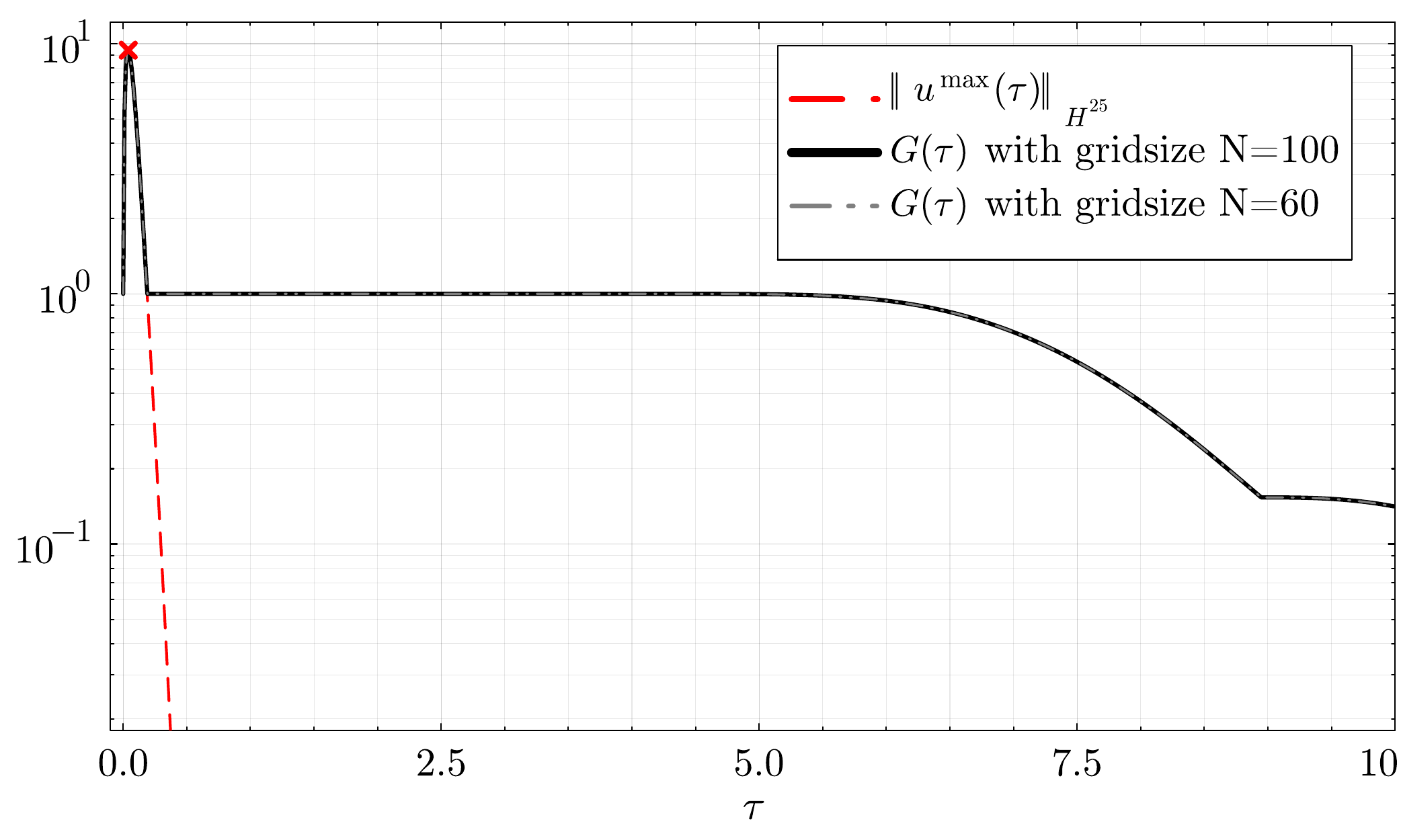}}
\caption{\textbf{Left:} Optimal perturbation and energy growth curve $G_W(\tau)$ (black dash) for complex charged scalar QNMs of the RN-AdS$_4$ black brane with $M=10$. $E_\psi$ (solid black curve) is shown to transiently grow before modally decaying at asymptotic time. The additional energy is borrowed from the energy bath $E_F$ via a transient form of superradiance, as can be seen from the first correction to $E_F$, $E_{F}^{(2)}$ (solid blue curve). The example shown corresponds to the probe limit $q\to\infty$ with $\mu q=3.9$, spatial momentum $\vec{k}=0$, with target time $\tau_*=2.7$. Figure taken from \cite{Carballo:2025ajx}. \textbf{Right:} Growth curve $G$ and the norm of $u^\mathrm{max}(\tau,x)$ obtained by time-evolving the optimal perturbation $u_{\mathrm{opt.}}(\tau_*=\tau_\mathrm{max},x)$ in the Sobolev $H^{25}$ norm. A transient growth is observed and yields a peak at $\tau_\mathrm{max}\approx \frac{1}{25}=0.04$, before a modal oscillatory decay. The growth curve $G$ is computed for two different resolutions $N=60$ (dashed gray curve) and $N=100$ (solid black curve), thus illustrating the convergence of the profile we observe on this panel. Figure taken from \cite{BessonV2}.}
    \label{fig:growth}    
\end{figure}

Building on \cite{Carballo:2024kbk}, \cite{Carballo:2025ajx} established the first case of transient energy growth in linear black hole perturbations considering RN-AdS$_4$ black branes at chemical potential $\mu$ linearly perturbed by a complex scalar $\psi$ with charge $q$. The key difference here is that the total energy of the system does not correspond to the energy of $\psi$ alone. In particular, in the $q\to\infty$ limit suppressing backreaction to the metric, $E$ receives contributions from both the scalar, $\psi$, and the gauge field, $A$,
\be
E=E_\psi+E_F,
\label{eq:total_energy}
\ee
where $F=dA$, which are coupled to each other through $q$. Then, choosing $\|\cdot\|_{E_\psi}$ to construct optimal perturbations in the same fashion as before, $E_\psi$ was shown to exhibit significant transient growth before asymptotic decay via a transient form of superradiance -- borrowing from the energy bath $E_F$ -- in the modally stable regime.\footnote{In AdS/CFT, this model is known as the holographic superconductor \cite{Gubser:2008px, Hartnoll:2008vx, Hartnoll:2008kx}, and it is linearly unstable for $T<T_c$ (or equivalently $\mu>\mu_c$) corresponding to the superconducting phase.} This is shown in the left panel of figure \ref{fig:growth}, which displays $G_W$ and $E_\psi$ for an optimal perturbation with $M=10$ QNMs exhibiting transient growth. The first correction to the background gauge field energy, which appears at quadratic order in perturbations, $E_F^{(2)}$, takes negative values implying transfer of energy from $A$ to $\psi$, while the total energy $E=E_\psi+E_F^{(2)}$ can only decrease due to losses to the horizon. Empirically, it was observed that the peak of the growth curve increases with $M$ within the range of values considered. Note that this is not a special feature of the $q\to\infty$ limit, the finite $q$ case is also examined in \cite{Carballo:2025ajx} with same qualitative results.

Transient behaviour has also been seen in Sobolev $H^p$ norms \eqref{sobolevdef} in \cite{BessonV2} in the P\"oschl-Teller toy model, corresponding to the Klein-Gordon equation in the static patch of de Sitter spacetime. Following a similar approach to \cite{Carballo:2024kbk,Carballo:2025ajx}, optimal perturbations $u_{\text{opt.}}(0,x)$ were obtained using a `generalised' singular value decomposition of the finite rank approximant of the evolution operator $e^{-i\mathcal{H}\tau}$, but this time without relying on a subspace of solutions $W$. For a target time $\tau_*$, these optimal perturbations maximise the Sobolev $H^p$ inner product such that $\langle u_{\text{opt.}}(\tau_*,x),u_{\text{opt.}}(\tau_*,x) \rangle_{H^p}=G(\tau_*)$.

In the case of $H^0$ norm (corresponding to the energy norm), no transient growth is observed. Similar to \cite{Carballo:2024kbk}, non-modal behaviour manifests itself as an initial transient plateau in $G$, followed by the expected modal decay, with a scalar field profile localised near the boundaries. However, unlike \cite{Carballo:2024kbk}, the results are not convergent as the duration of the plateau scales as $\log N$ with the number of points $N$ used in the numerical approximation, further motivating the use of $H^p$ norms with $p>0$.

In the case of $H^p$ Sobolev norm with $p>0$, transient growth is observed. Specifically, one obtains an initial `peak', which is followed by modal decay according to the lowest-lying QNM at late times; the right panel of figure \ref{fig:growth} exemplifies this behaviour for $p=25$, showing $G$ and $\langle u_{\text{opt.}}(\tau_*,x),u_{\text{opt.}}(\tau_*,x)\rangle_{H^{25}}$ for $\tau_*=\tau_{\mathrm{max}}$ corresponding to the time of the peak. Note that $u_{\text{opt.}}(\tau_*,x)$ is an order-$p$ polynomial. The profile of the optimal perturbation $u_{\text{opt.}}(\tau,x)$ is found to be numerically convergent and, importantly, it resides in the bulk of the geometry; this is different to the energy-norm case where the optimal perturbations was peaked near the boundaries. Applying the `Keldysh' spectral decomposition scheme to $u_{\text{opt.}}(0,x)$ shows that most of the transient peak originates from the $(p+1)$-th pair of QNMs, ordered by decreasing $\im{\omega}$; note that the decay rate of these modes is $\frac{1}{p}$.

As the order $p$ of the $H^p$ Sobolev norm is increased, the peak in the growth curve increase as $G(\tau_{\mathrm{max}})=\max_{\tau\geq 0}G(\tau)\sim p$ and moves to shorter timescales, $\tau_{\mathrm{max}}\sim 1/p$. The scaling of $\tau_{\mathrm{max}}$ is a result of the decay rate of the QNM giving rise to the majority of the transient peak mentioned above. 

Lastly, it is illuminating to understand the existence of $H^p$-transient growth in the context of energy conservation. Specifically, the $H^p$-norm satisfies
\bea
\label{eq:total_energy_Hp}
E[u]=\langle u,u\rangle_{H^p}- \sum_{j=1}^p \left\Vert{ 
    \partial_x^j u}\right\Vert_{E}^2 \ ,
\eea
where $E[u]$ is conserved up to boundary terms. In a way analogous to \ref{eq:total_energy}, $H^p$-transient growth is permitted as a result of transfer of weight between the two terms in the right hand side of \ref{eq:total_energy_Hp}.

Let us conclude this section with a comparison of the two methods discussed above, truncating the set of QNMs or using higher-derivative norms. Both approaches provide a way of regulating the UV and are equally easy to implement. The motivations for using them are different: in the former case the motivation was a physical truncation of the theory to low energy modes inspired by analogous constructions in hydrodynamics, while in the latter case the motivation was a consideration of regularity. The truncation method results in a finite dimensional Hilbert space which can be convenient to work with. The physical interpretation of the $H^p$-norm remains an open question.

\section{Discussion}

This short review summarises recent work on transient phenomena in black hole dynamics. The lack of normality of the evolution operator, emerging as a consequence of the dissipative nature of black hole spacetimes, results in the non-orthogonality of QNMs. This, in turn, allows for linear perturbations to exhibit non-modal behaviour (either in the form of transient growth or lack of decay) before eventually conforming to modal decay.

The existence of transients can be inferred from frequency-domain computations involving the pseudospectrum: the protrusion of pseudospectral contour lines in the unstable half plane indicates an unstable perturbed spectrum, and hence non-modal behaviour. In order to observe transient growth, the protrusion needs to be larger than the size of the external perturbation $\epsilon$, giving rise to a Kreiss constant $\mathcal{K}>1$. This raises again the issue of the numerical convergence of the pseudospectrum as discussed in section \ref{section2}, and motivates the exploration of the truncated-Hamiltonian pseudospectrum of \cite{Carballo:2025ajx}.

Time-domain results exhibit striking qualitative similarities to the prototypical example of transient effects in the transition to turbulence in Navier-Stokes shear flows. Two particularly interesting questions that currently remain open relate to the non-linear evolution sourced by such initial data and the potential connection with the Aretakis instability.

Black hole QNMs have been a central focus of gravitational physics for over half a century, yet it remains striking that we still lack a full understanding of the consequences stemming from the absence of a spectral theorem in this context. This gap points to an exciting new direction in the field, suggesting that much remains to be uncovered. Particularly compelling questions include how much of the gravitational wave signal emanating from a binary merger can be attributed to linear transient dynamics, as well as the role of transients in strongly coupled systems, such as the quark-gluon plasma and high-temperature superconductors, via the AdS/CFT correspondence. Other arenas include analogue gravity systems, where fluid or optical setups mimic aspects of black hole spacetimes.

\section*{Acknowledgments}
The authors would like to thank José Luis Jaramillo and Frans Pretorius for discussions. J.B. is supported by the project QuanTEdu-France 22-CMAS-0001. J.C. is supported by the Royal Society Research Grant RF\textbackslash ERE \textbackslash 210267. 
C.P. is supported by a Royal Society -- Research Ireland University Research Fellowship via grant URF\textbackslash R1\textbackslash 211027. 
B.W. is supported by a Royal Society University Research Fellowship URF\textbackslash R\textbackslash 231002 and in part by the STFC consolidated grant ST/T000775/1.

\bibliographystyle{Frontiers-Vancouver}
\bibliography{refs}

\begin{thebibliography}{38}
\expandafter\ifx\csname natexlab\endcsname\relax\def\natexlab#1{#1}\fi
\expandafter\ifx\csname urlstyle\endcsname\relax
  \expandafter\ifx\csname doi\endcsname\relax
  \def\doi#1{doi:\discretionary{}{}{}#1}\fi \else
  \expandafter\ifx\csname doi\endcsname\relax
  \def\doi{doi:\discretionary{}{}{}\begingroup \urlstyle{rm}\Url}\fi \fi
\expandafter\ifx\csname selectlanguage\endcsname\relax
  \def\selectlanguage#1{}\fi

\bibitem[{Berti et~al.(2009)Berti, Cardoso, and Starinets}]{Berti:2009kk}
Berti E, Cardoso V, Starinets AO.
\newblock {Quasinormal modes of black holes and black branes}.
\newblock {\em Class. Quant. Grav.\/} {\bf 26} (2009) 163001.
\newblock \doi{10.1088/0264-9381/26/16/163001}.

\bibitem[{Konoplya and Zhidenko(2011)}]{Konoplya:2011qq}
Konoplya RA, Zhidenko A.
\newblock {Quasinormal modes of black holes: From astrophysics to string
  theory}.
\newblock {\em Rev. Mod. Phys.\/} {\bf 83} (2011) 793--836.
\newblock \doi{10.1103/RevModPhys.83.793}.

\bibitem[{Lagos and Hui(2023)}]{Lagos:2022otp}
Lagos M, Hui L.
\newblock {Generation and propagation of nonlinear quasinormal modes of a
  Schwarzschild black hole}.
\newblock {\em Phys. Rev. D\/} {\bf 107} (2023) 044040.
\newblock \doi{10.1103/PhysRevD.107.044040}.

\bibitem[{Pantelidou and Withers(2023)}]{Pantelidou:2022ftm}
Pantelidou C, Withers B.
\newblock {Thermal three-point functions from holographic Schwinger-Keldysh
  contours}.
\newblock {\em JHEP\/} {\bf 04} (2023) 050.
\newblock \doi{10.1007/JHEP04(2023)050}.

\bibitem[{Policastro et~al.(2001)Policastro, Son, and
  Starinets}]{Policastro:2001yc}
Policastro G, Son DT, Starinets AO.
\newblock {The Shear viscosity of strongly coupled N=4 supersymmetric
  Yang-Mills plasma}.
\newblock {\em Phys. Rev. Lett.\/} {\bf 87} (2001) 081601.
\newblock \doi{10.1103/PhysRevLett.87.081601}.

\bibitem[{Kovtun and Starinets(2005)}]{Kovtun:2005ev}
Kovtun PK, Starinets AO.
\newblock {Quasinormal modes and holography}.
\newblock {\em Phys. Rev. D\/} {\bf 72} (2005) 086009.
\newblock \doi{10.1103/PhysRevD.72.086009}.

\bibitem[{Baibhav et~al.(2023)Baibhav, Cheung, Berti, Cardoso, Carullo, Cotesta
  et~al.}]{Baibhav:2023clw}
Baibhav V, Cheung MHY, Berti E, Cardoso V, Carullo G, Cotesta R, et~al.
\newblock {Agnostic black hole spectroscopy: Quasinormal mode content of
  numerical relativity waveforms and limits of validity of linear perturbation
  theory}.
\newblock {\em Phys. Rev. D\/} {\bf 108} (2023) 104020.
\newblock \doi{10.1103/PhysRevD.108.104020}.

\bibitem[{York(1983)}]{York:1983zb}
York JW Jr.
\newblock {Dynamical Origin of Black Hole Radiance}.
\newblock {\em Phys. Rev. D\/} {\bf 28} (1983) 2929.
\newblock \doi{10.1103/PhysRevD.28.2929}.

\bibitem[{Hintz and Vasy(2017)}]{Hintz:2015jkj}
Hintz P, Vasy A.
\newblock {Analysis of linear waves near the Cauchy horizon of cosmological
  black holes}.
\newblock {\em J. Math. Phys.\/} {\bf 58} (2017) 081509.
\newblock \doi{10.1063/1.4996575}.

\bibitem[{Jafferis et~al.(2015)Jafferis, Lupsasca, Lysov, Ng, and
  Strominger}]{Jafferis:2013qia}
Jafferis DL, Lupsasca A, Lysov V, Ng GS, Strominger A.
\newblock {Quasinormal quantization in de Sitter spacetime}.
\newblock {\em JHEP\/} {\bf 01} (2015) 004.
\newblock \doi{10.1007/JHEP01(2015)004}.

\bibitem[{Green et~al.(2023)Green, Hollands, Sberna, Toomani, and
  Zimmerman}]{Green:2022htq}
Green SR, Hollands S, Sberna L, Toomani V, Zimmerman P.
\newblock {Conserved currents for a Kerr black hole and orthogonality of
  quasinormal modes}.
\newblock {\em Phys. Rev. D\/} {\bf 107} (2023) 064030.
\newblock \doi{10.1103/PhysRevD.107.064030}.

\bibitem[{London(2023)}]{London:2023aeo}
London LT.
\newblock {A radial scalar product for Kerr quasinormal modes}.
\newblock {\em arXiv:2312.17678\/}  (2023).

\bibitem[{Arnaudo et~al.(2025)Arnaudo, Carballo, and Withers}]{Arnaudo:2025bnm}
Arnaudo P, Carballo J, Withers B.
\newblock {QNM orthogonality relations for AdS black holes}.
\newblock {\em arXiv:2505.04696\/}  (2025).

\bibitem[{Krejcirik et~al.(2015)Krejcirik, Siegl, Tater, and
  Viola}]{Krejcirik:2014kaa}
Krejcirik D, Siegl P, Tater M, Viola J.
\newblock {Pseudospectra in non-Hermitian quantum mechanics}.
\newblock {\em J. Math. Phys.\/} {\bf 56} (2015) 103513.
\newblock \doi{10.1063/1.4934378}.

\bibitem[{Trefethen and Embree(2005)}]{trefethen2005spectra}
Trefethen L, Embree M.
\newblock {\em Spectra and Pseudospectra: The Behavior of Nonnormal Matrices
  and Operators\/} (Princeton University Press) (2005).

\bibitem[{Gasperin and Jaramillo(2022)}]{Gasperin:2021kfv}
Gasperin E, Jaramillo JL.
\newblock {Energy scales and black hole pseudospectra: the structural role of
  the scalar product}.
\newblock {\em Class. Quant. Grav.\/} {\bf 39} (2022) 115010.
\newblock \doi{10.1088/1361-6382/ac5054}.

\bibitem[{Jaramillo et~al.(2021)Jaramillo, Panosso~Macedo, and
  Al~Sheikh}]{Jaramillo:2020tuu}
Jaramillo JL, Panosso~Macedo R, Al~Sheikh L.
\newblock {Pseudospectrum and Black Hole Quasinormal Mode Instability}.
\newblock {\em Phys. Rev. X\/} {\bf 11} (2021) 031003.
\newblock \doi{10.1103/PhysRevX.11.031003}.

\bibitem[{Nollert and Price(1999)}]{Nollert:1998ys}
Nollert HP, Price RH.
\newblock {Quantifying excitations of quasinormal mode systems}.
\newblock {\em J. Math. Phys.\/} {\bf 40} (1999) 980--1010.
\newblock \doi{10.1063/1.532698}.

\bibitem[{Nollert(1996)}]{Nollert:1996rf}
Nollert HP.
\newblock {About the significance of quasinormal modes of black holes}.
\newblock {\em Phys. Rev. D\/} {\bf 53} (1996) 4397--4402.
\newblock \doi{10.1103/PhysRevD.53.4397}.

\bibitem[{Jaramillo(2022)}]{Jaramillo:2022kuv}
Jaramillo JL.
\newblock {Pseudospectrum and binary black hole merger transients}.
\newblock {\em Class. Quant. Grav.\/} {\bf 39} (2022) 217002.
\newblock \doi{10.1088/1361-6382/ac8ddc}.

\bibitem[{Boyanov et~al.(2023)Boyanov, Destounis, Panosso~Macedo, Cardoso, and
  Jaramillo}]{Boyanov:2022ark}
Boyanov V, Destounis K, Panosso~Macedo R, Cardoso V, Jaramillo JL.
\newblock {Pseudospectrum of horizonless compact objects: A bootstrap
  instability mechanism}.
\newblock {\em Phys. Rev. D\/} {\bf 107} (2023) 064012.
\newblock \doi{10.1103/PhysRevD.107.064012}.

\bibitem[{Carballo and Withers(2024)}]{Carballo:2024kbk}
Carballo J, Withers B.
\newblock {Transient dynamics of quasinormal mode sums}.
\newblock {\em JHEP\/} {\bf 10} (2024) 084.
\newblock \doi{10.1007/JHEP10(2024)084}.

\bibitem[{Chen et~al.(2024)Chen, Wu, and Guo}]{Chen:2024mon}
Chen JN, Wu LB, Guo ZK.
\newblock {The pseudospectrum and transient of Kaluza\textendash{}Klein black
  holes in Einstein\textendash{}Gauss\textendash{}Bonnet gravity}.
\newblock {\em Class. Quant. Grav.\/} {\bf 41} (2024) 235015.
\newblock \doi{10.1088/1361-6382/ad89a1}.

\bibitem[{Boyanov et~al.(2024)Boyanov, Cardoso, Destounis, Jaramillo, and
  Macedo}]{Boyanov:2024}
Boyanov V, Cardoso V, Destounis K, Jaramillo JL, Macedo RP.
\newblock Structural aspects of the anti--de sitter black hole pseudospectrum.
\newblock {\em Phys. Rev. D\/} {\bf 109} (2024) 064068.
\newblock \doi{10.1103/PhysRevD.109.064068}.

\bibitem[{Warnick(2024)}]{warnick2024stability}
Warnick C.
\newblock (in) stability of de sitter quasinormal mode spectra.
\newblock {\em arXiv:2407.19850\/}  (2024).

\bibitem[{Besson and Jaramillo(2025)}]{BessonV2}
Besson J, Jaramillo JL.
\newblock {Quasi-normal mode expansions of black hole perturbations: a
  hyperboloidal Keldysh's approach}.
\newblock {\em arXiv:2412.02793v2\/}  (2025).

\bibitem[{Warnick(2015)}]{Warnick:2013hba}
Warnick CM.
\newblock {On quasinormal modes of asymptotically anti-de Sitter black holes}.
\newblock {\em Commun. Math. Phys.\/} {\bf 333} (2015) 959--1035.
\newblock \doi{10.1007/s00220-014-2171-1}.

\bibitem[{Bizo\'n et~al.(2020)Bizo\'n, Chmaj, and Mach}]{Bizon:2020qnd}
Bizo\'n P, Chmaj T, Mach P.
\newblock {A toy model of hyperboloidal approach to quasinormal modes}.
\newblock {\em Acta Phys. Polon. B\/} {\bf 51} (2020) 1007.
\newblock \doi{10.5506/APhysPolB.51.1007}.

\bibitem[{Carballo et~al.(2025)Carballo, Pantelidou, and
  Withers}]{Carballo:2025ajx}
Carballo J, Pantelidou C, Withers B.
\newblock {Non-modal effects in black hole perturbation theory: Transient
  Superradiance}.
\newblock {\em arXiv:2503.05871\/}  (2025).

\bibitem[{Reddy et~al.(1993)Reddy, Schmid, and Henningson}]{Reddy93}
Reddy SC, Schmid PJ, Henningson DS.
\newblock {Pseudospectra of the Orr-Sommerfeld Operator}.
\newblock {\em SIAM Journal on Applied Mathematics\/} {\bf 53} (1993) 15--47.

\bibitem[{Gustavsson(1991)}]{Gustavsson_1991}
Gustavsson LH.
\newblock {Energy growth of three-dimensional disturbances in plane Poiseuille
  flow}.
\newblock {\em Journal of Fluid Mechanics\/} {\bf 224} (1991) 241–260.
\newblock \doi{10.1017/S002211209100174X}.

\bibitem[{Henningson et~al.(1993)Henningson, Lundbladh, and
  Johansson}]{Henningson_Lundbladh_Johansson_1993}
Henningson DS, Lundbladh A, Johansson AV.
\newblock A mechanism for bypass transition from localized disturbances in
  wall-bounded shear flows.
\newblock {\em Journal of Fluid Mechanics\/} {\bf 250} (1993) 169–207.
\newblock \doi{10.1017/S0022112093001429}.

\bibitem[{Butler and Farrell(1992)}]{ButlerFarrell}
Butler KM, Farrell BF.
\newblock {Three‐dimensional optimal perturbations in viscous shear flow}.
\newblock {\em Physics of Fluids A: Fluid Dynamics\/} {\bf 4} (1992)
  1637--1650.
\newblock \doi{10.1063/1.858386}.

\bibitem[{Reddy and Henningson(1993)}]{Reddy_Henningson_1993}
Reddy SC, Henningson DS.
\newblock Energy growth in viscous channel flows.
\newblock {\em Journal of Fluid Mechanics\/} {\bf 252} (1993) 209–238.
\newblock \doi{10.1017/S0022112093003738}.

\bibitem[{Trefethen et~al.(1993)Trefethen, Trefethen, Reddy, and
  Driscoll}]{Trefethen1993}
Trefethen LN, Trefethen AE, Reddy SC, Driscoll TA.
\newblock Hydrodynamic stability without eigenvalues.
\newblock {\em Science\/} {\bf 261} (1993) 578--584.
\newblock \doi{10.1126/science.261.5121.578}.

\bibitem[{Gubser(2008)}]{Gubser:2008px}
Gubser SS.
\newblock {Breaking an Abelian gauge symmetry near a black hole horizon}.
\newblock {\em Phys. Rev. D\/} {\bf 78} (2008) 065034.
\newblock \doi{10.1103/PhysRevD.78.065034}.

\bibitem[{Hartnoll et~al.(2008{\natexlab{a}})Hartnoll, Herzog, and
  Horowitz}]{Hartnoll:2008vx}
Hartnoll SA, Herzog CP, Horowitz GT.
\newblock {Building a Holographic Superconductor}.
\newblock {\em Phys. Rev. Lett.\/} {\bf 101} (2008{\natexlab{a}}) 031601.
\newblock \doi{10.1103/PhysRevLett.101.031601}.

\bibitem[{Hartnoll et~al.(2008{\natexlab{b}})Hartnoll, Herzog, and
  Horowitz}]{Hartnoll:2008kx}
Hartnoll SA, Herzog CP, Horowitz GT.
\newblock {Holographic Superconductors}.
\newblock {\em JHEP\/} {\bf 12} (2008{\natexlab{b}}) 015.
\newblock \doi{10.1088/1126-6708/2008/12/015}.

\end{thebibliography}

\end{document}